\documentclass[]{aastex631}

\usepackage[utf8]{inputenc}
\usepackage{amsmath}
\usepackage{graphicx}
\usepackage{subfigure}
\usepackage{amssymb}
\usepackage{amsthm}
\usepackage{subfigure}
\usepackage{xcolor}
\usepackage[figuresright]{rotating}
\usepackage{txfonts}
\usepackage{CJKutf8}

\newcommand       \Teff         {T_{\rm {eff}}}
\newcommand       \GBp          {G_{\rm BP}}

\newcommand       \CBpRp        {(G_{\rm BP}-G_{\rm RP})}
\newcommand       \FeH          {\rm \left[Fe/H \right]}
\newcommand       \lgg         {{\rm \log}\ g}

\shorttitle{Estimation of OB stellar fluxes observable by FAST and SKA}
\shortauthors{Huang et al.}

\graphicspath{{./}{figures/}}

\begin{document}
\begin{CJK}{UTF8}{gbsn}

\title{Estimation of the flux at 1450MHz of OB stars for FAST and SKA}

\correspondingauthor{Biwei Jiang}
\email{bjiang@bnu.edu.cn}

\author[0000-0002-4046-2344]{Qichen Huang}
\affiliation{Institute for Frontiers in Astronomy and Astrophysics,
            Beijing Normal University,  Beijing 102206, China}
\affiliation{Department of Astronomy, Beijing Normal University, Beijing 100875, Peopleʼs Republic of China}
\email{202121160007@mail.bnu.edu.cn}

\author[0000-0003-3168-2617]{Biwei Jiang}
\affiliation{Institute for Frontiers in Astronomy and Astrophysics,
            Beijing Normal University,  Beijing 102206, China}
\affiliation{Department of Astronomy, Beijing Normal University, Beijing 100875, Peopleʼs Republic of China}

\author[0000-0003-0777-7392]{Dingshan Deng}
\affiliation{Lunar and Planetary Laboratory, The University of Arizona, Tucson, AZ 85721, USA}

\author[0000-0003-1178-5176]{Bin Yu}
\affiliation{Songshan Lake Future School, Dongguan 523808, Peopleʼs Republic of China}

\author[0000-0002-3171-5469]{Albert Zijlstra}
\affiliation{Department of Physics and Astronomy, The University of Manchester, Manchester M13 9PL, UK}

\begin{abstract}
Radio observation is crucial to understanding the wind mechanism of OB stars but very scarce. This work estimates the flux at 1450MHz ($S_{\rm 1.4GHz}$) of about 5,000 OB stars identified by the LAMOST spectroscopic survey and confirmed by the Gaia astrometric as well as astrophysical measurements. The calculation is performed under the free-free emission mechanism for wind with the mass loss rate derived from stellar parameters. The estimated $S_{\rm 1.4GHz}$ distributes from $10^{-11}$Jy to $10^{-3}$Jy with the peak at about $10^{-8}$Jy. This implies that the complete SKA-II can detect more than half of them, and some tens of objects are detectable by FAST without considering source confusion. An array of FAST would increase the detectable sample by two orders of magnitude.
\end{abstract}

\keywords{OB stars(1141) --- Radio continuum (1340) --- Stellar winds(1636) --- Radio telescopes(1360)}


\section{Introduction} \label{sec:intro}

The massive OB stars usually have strong winds that produce significant mass loss rate on the order of $10^{-4}-10^{-6} {\rm M_{\odot} yr^{-1}}$, which greatly influences the evolution of the star and the surrounding interstellar medium \citep{matthews2018}. Thanks to \cite{castor1975} who described the theory of radiation-driven wind, we now have a basic understanding of how this kind of stellar wind is generated. However, there are still some important questions which radio observations can uniquely answer. First, in the strong stellar wind of OB stars, the dominant radio emission is considered to be the thermal free-free radiation from the ionized gas around the star. A big advantage of using radio flux to determine the mass loss rate is that its radiation is generated farther away in the stellar wind than UV or X-ray radiation and can be considered to have almost reached its terminal velocity. Consequently, it has no strong dependence on the ionization environment, internal velocity field, etc., and the mass loss rate can be determined more precisely. Moreover,  stellar wind is dynamically unstable for OB stars, which leads to shock formation and clumping. This clumping will enhance the radio emission and produce variation in the spectral index as a function of frequency \citep{daley2016}. The radio observation therefore provides unique insights into the nature of this stellar wind.

Radio observations of OB stars have been conducted by some groups. \citet{isequilla2019} observed the  Cygnus OB2 and OB8 regions by the Giant Metrewave Radio Telescope (GMRT), and detected nine young massive stars that have the flux at 325MHz and 610MHz above 2 mJy. Previously, \citet{kennedy2010} examined fifty observations across frequencies between 1.4 GHz and 43 GHz using VLA to observe the Cygnus OB2 No.5 system. As a binary system with an OB star orbiting, its radio flux per observation exceeded 2.5 mJy. Their model derives a mass loss rate of $3.4 \times 10^{-5}{\rm M_\odot yr^{-1}}$ for Cyg OB2 No. 5, which is unusually high for an Of supergiant and comparable to that of WR stars. On the other hand, the e-Merlin Cyg OB2 Radio Survey (COBRaS) project has accumulated about 300 hours of observations on 1,000 OB stars in the Cygnus OB2 association at 5GHz and 1.6GHz \citep{willis2011}. They reported that the 21cm flux of O3 to O6 supergiant and giant stars are less than $\sim 70 {\rm \mu}$Jy, which may imply an upper limit of mass loss rate of $\sim 4.5 \times 10^{-6}{\rm M_\odot yr^{-1}}$ for the O3 supergiants  and $\leq 2.9 \times 10^{-6}{\rm M_\odot yr^{-1}}$ for B0 to B1 supergiants.

The detection of OB star radio emission is generally difficult due to its weakness. With the development of large radio telescope like FAST \citep{jiang2020} and the planned big project such as SKA \citep{combes2015}, the sensitivity is greatly increased and the detection of large number of OB stars becomes possible, consequently stellar wind mechanism and properties can be better understood. Recently, \cite{yu2021}  generated a star catalog that includes the OB stars through the Besançon model, and used this as a target to estimate the observability of the SKA telescope. They reported that the SKA can detect 1500 Be stars and 50 OB stars per square degree out to several kpc at the limit of 10 nJy at 5GHz.

The FAST telescope is a pointing facility that cannot compare with the wide FOV of SKA. In this regard, the catalog generated by the Besançon model cannot offer the positions of the real targets. The five-hundred-meter aperture in combination with the receiver's system temperature $<30$K of FAST makes it one of the most sensitive telescopes at the decimeter wavebands currently available. The estimation of the radio flux of OB stars would provide suitable targets for studying the ionized wind. Synchronous with the development of radio astronomy, the sample of OB stars is significantly augmented by large-scale optical spectroscopic surveys.  The recent publication of the new OB star catalog by \citet{guo2021} includes 16,032 early-type stars through measuring the equivalent widths of several absorption line profiles
using low-resolution spectra from the LAMOST database.  In addition, the release of Gaia/DR3 data has brought us high-precision distance as well as stellar parameters and interstellar extinction. These data lead to the feasibility of predicting the detectability of OB stars by the large radio facilities based on the real stars instead of the model ones. It will also be helpful to identify the radio stars in the SKA large area observations.

%
%
%

This work intends to estimate the radio flux of a large sample of OB stars and then to predict the observability by the FAST and SKA telescopes. In brief, the initial sample of OB stars is taken from the catalog identified from the LAMOST optical spectrum. Being a reflective
Schmidt telescope with a diameter of 5-m and an FOV of 5\degr squared, LAMOST can provide the spectra of about 4000 objects in one exposure and has accumulated the spectra of nearly 10 million stars \citep{cui2012, luo2015}. This sample is further cleaned by the Gaia measurements. With the stellar parameters derived from Gaia, the mass loss rate and the radio flux at 1450MHz are estimated, to be compared with the visibility and sensitivity of the FAST and SKA telescopes in order to predict the number of observable objects. The data is presented in Section \ref{sec:The Sample}, the method in Section \ref{sec:calculation}, and the result in Section \ref{sec:Radio flux}.

\section{The Sample} \label{sec:The Sample}

The initial sample is the OB star catalog of \citet{guo2021}, who used the Stellar LAbel Machine (SLAM), which is a machine learning method, to screen the spectra taken by LAMOST.  They identified 578 and 3931 OB stars within the training parameters from the LAMOST medium-resolution spectra (MRS) and low-resolution spectra (LRS) respectively, and calculated their effective temperature $\Teff$, surface gravity $\lgg$, metal abundance $\FeH$, etc. For the stars outside the parameters' range, they calculated the atmospheric parameters by extrapolation from SLAM. Adding this part from extrapolation, \cite{guo2021} built up a catalog of 9287 and 22292 OB stars from LAMOST/MRS and LRS respectively. After removing the duplicate sources while keeping the result from the highest SNR spectrum, the total number of sources becomes 21731. This catalog is the first one that consists of consistently derived stellar labels for such a large sample of early-type stars.\citep{guo2021}.

This initial sample is further cleaned by using the Gaia/DR3 data. For hundreds of millions of stars, Gaia/DR3 provides  stellar parameters with a method called the General Stellar Parameterizer from Photometry (GSP-Phot), including effective temperature, surface gravity, metallicity, absolute magnitude, radius, distance, and extinction,  with the information from astrometry, photometric and low-resolution BP/RP spectra \citep{andrae2022}.  We first look up the location of the initial sample stars in the color-magnitude diagram to confirm that the object is an early-type star.  The extinction is taken from the Gaia extinction measurements through Gaia's low-resolution BP/RP spectroscopy, which have been validated in several places, such as the Sun-like stars project of \cite{creevey2022}. \citet{delchambre2022} proved that the overall extinction map of the Milky Way drawn by Gaia/DR3 is in good agreement with the Planck data  and  with the SFD98 2D extinction map \citep{schlegel1998}. In this step, the sources whose extinction is unavailable in the Gaia/DR3 catalog are rejected, which leaves 13,729 sources in the sample. For the distance,  the geometric distance of Gaia/EDR3 given by \cite{bailer2021} is adopted and available for all the sources.

\begin{figure}
    \centering
    \includegraphics[scale=0.45]{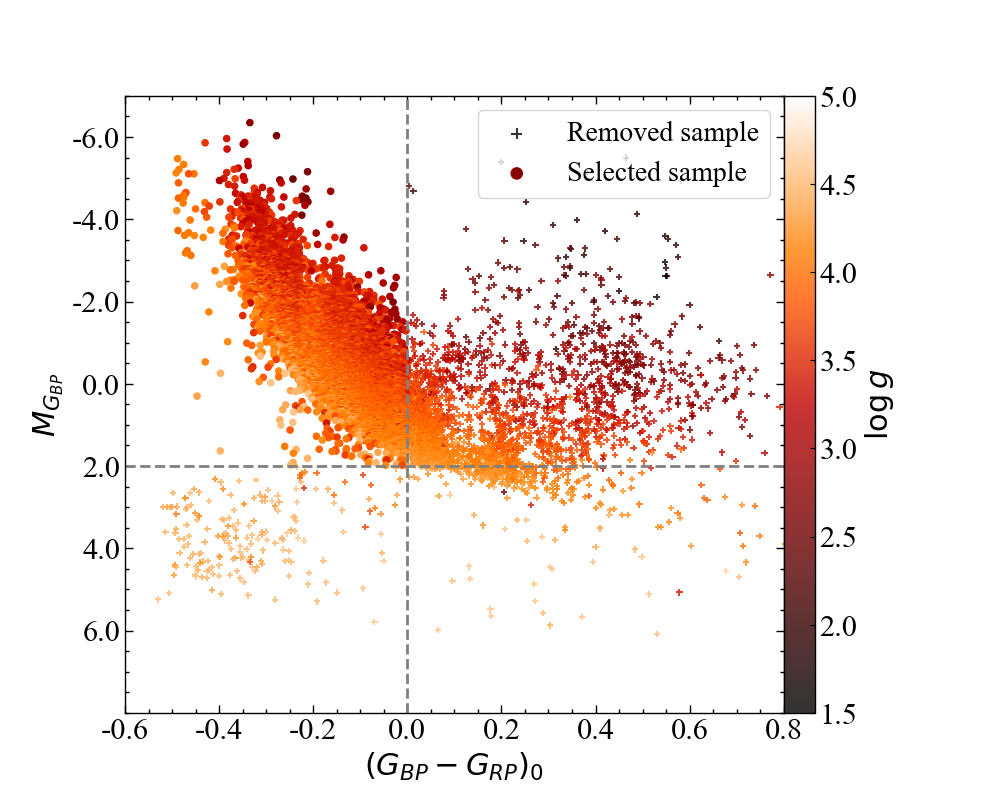}
    \caption{The color-magnitude diagram of all the initial sample stars, among which the star kept in the final sample is denoted by dot and the removed star is denoted by cross. The color bar indicates the surface gravity. }
    \label{fig:1}
\end{figure}

The color-magnitude diagram,  $\CBpRp_{0}$ vs. $M_{\rm G_{BP}}$, of the stars in the initial sample is presented in Figure \ref{fig:1}. The stars can be mainly divided into three groups. One group of stars has the absolute magnitude $M_{\rm G_{BP}}$ fainter than 2\,mag, which disagrees with the brightness of OB stars. Meanwhile, their intrinsic color index $\CBpRp_{0}$ is mostly $<0$, indicating a high temperature. These are probably hot subdwarf stars. The fact that their surface gravity denoted by the color in Figure \ref{fig:1} is generally $>4.5$  supports the subdwarf classification. Another group of stars has the intrinsic color index $\CBpRp_{0}>0.2$, which is apparently redder than the OB stars, and they may be giant stars considering their $M_{\rm G_{BP}}$ brighter than 0\,mag. Their relatively small surface gravity, $\lgg <2.5$ supports to classify them into giant stars. These two groups of stars are excluded. Consequently, the left sample contains 10646 stars which coincide with OB stars in color, brightness and surface gravity.


\section{Calculation of stellar parameters} \label{sec:calculation}

The estimation of radio flux needs several stellar parameters, including distance, luminosity, mass, effective temperature in combination with other parameters.

\subsection{Distance, Extinction and Effective Temperature}

\begin{figure}
    \centering
    \includegraphics[scale=0.45]{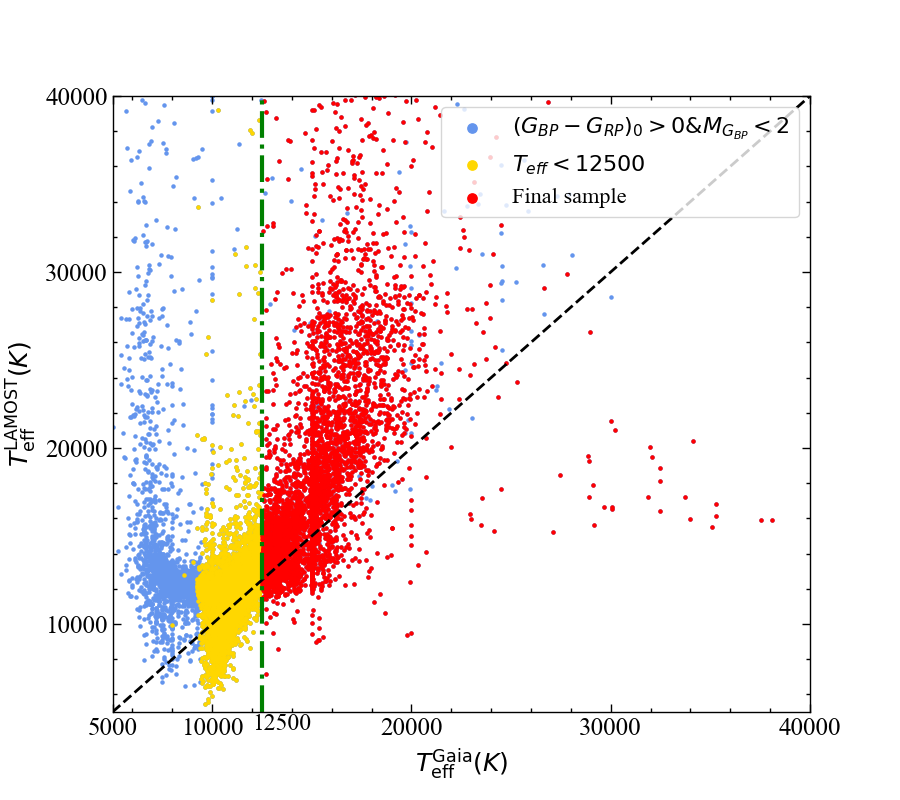}
    \caption{Comparison of $T_{\rm eff}$  derived by Gaia/DR3 and LAMOST. The blue dot denotes the objects we filtered out by $\CBpRp_{0} \textgreater 0$ and $M_{\rm G_{BP}} \textless 2$. The yellow dot denotes the objects that passes photometry screening but $T_{\rm eff}$ is lower than 12500K and therefore removed as well. The red dot denotes the stars for which we finally calculate the radio flux.}
    \label{fig:2}
\end{figure}

\begin{figure}
    \centering
    \subfigure[]{\includegraphics[width=8cm]{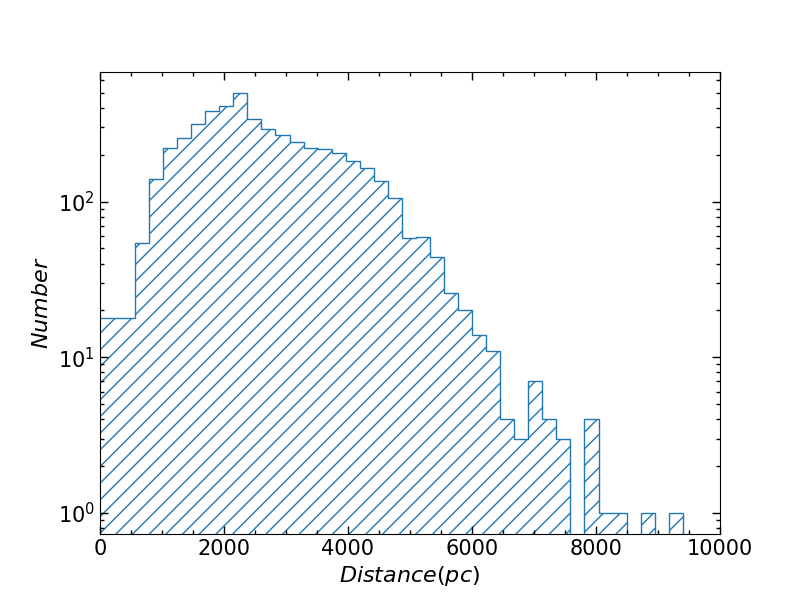}}
    \subfigure[]{\includegraphics[width=8cm]{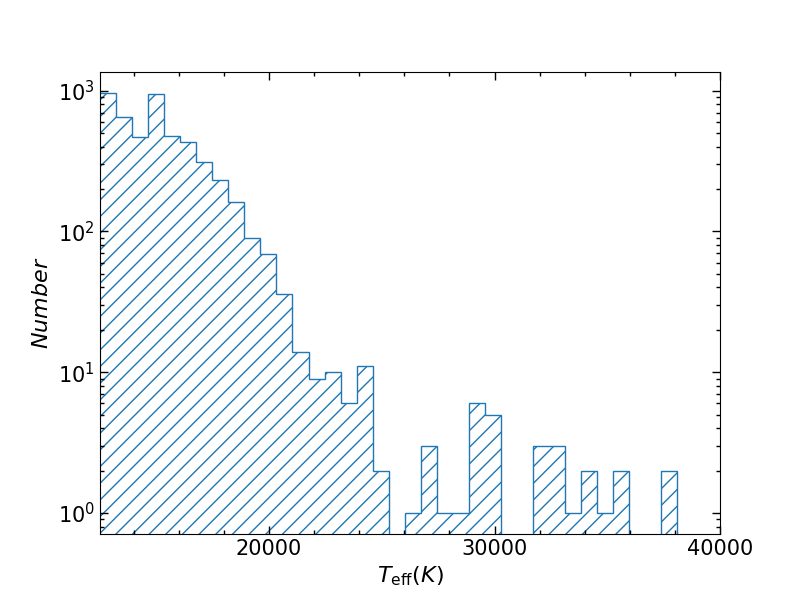}}
    \\ 
    \centering
    \subfigure[]{\includegraphics[width=8cm]{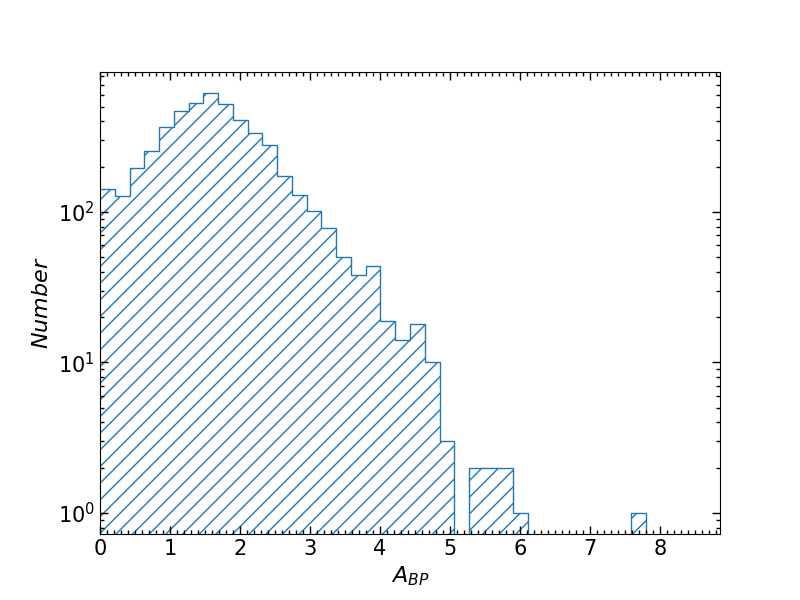}}
    \subfigure[]{\includegraphics[width=8cm]{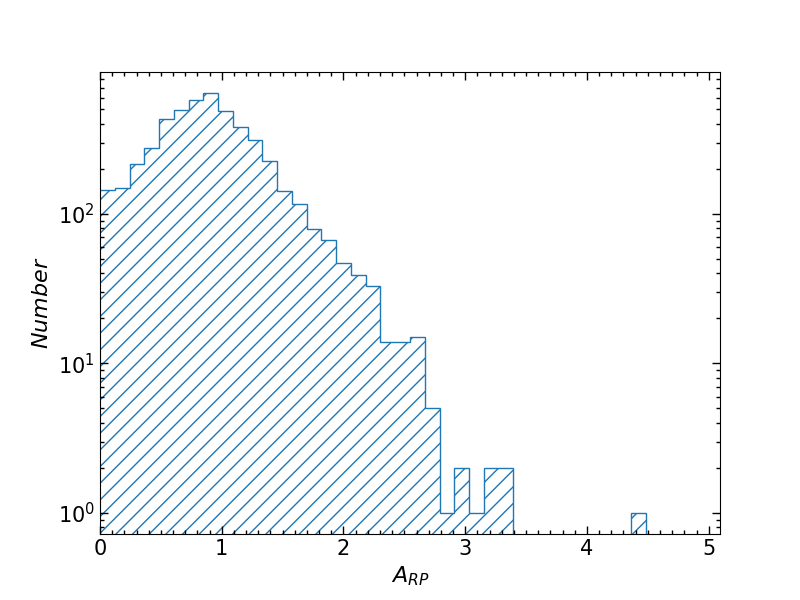}}
    \caption{Histogram of distance, effective temperature and interstellar extinction in the Gaia/BP and RP band of the sample stars.}
    \label{fig:03}
\end{figure}

As mentioned above, the distance is taken from the geometric distance of Gaia/EDR3 given by \cite{bailer2021}. On the effective temperature, both Gaia and LAMOST give the values. Gaia/GSP-Phot yielded very good results, with half of the values being within 170K of the literatures. In comparison with the effective temperature in the LAMOST catalog in Figure \ref{fig:2}, the value from GSP-Phot do not differ much from LAMOST in the range of 10000\,K to 15,000\,K, which is the most numerous part of the catalog. Meanwhile, the GSP-Phot temperature is significantly lower than the LAMOST one in the ranges of 5000K-10000K and 15000K to 20000K. Although the Gaia/DR3 temperature may be overestimated for some stars due to the overestimated extinction by GSP-Phot \citep{andrae2022}, we still choose the Gaia/DR3 temperature instead of the LAMOST temperature. This is because the Gaia/DR3 parameters are much more significantly consistent between the parameters  (effective temperature, surface gravity and absolute magnitude) than LAMOST. Furthermore, \citet{guo2021} is aware that the stellar parameters from the extrapolation are not highly reliable. This can be attributed to the additional constraint from the distance by Gaia. Consequently, 3915 stars in the initial LAMOST sample have the Gaia effective temperature smaller than 10000\,K and are dropped from the further analysis. Besides, the stars with $T_{\rm eff} \in 10000-12500$\,K are dropped as well because no appropriate calculation can be made for their mass loss rates to be described later.  Consistently, the extinction is retrieved from the Gaia/DR3 catalog. The histogram of these parameters are shown in Figure \ref{fig:03}.

%
%
%

\subsection{Luminosity and Mass}

The luminosity is calculated using the bolometric correction (BC) for the Gaia blue band $\GBp$. \cite{pedersen2020} derived the formulae to calculate the BC to the hot magnitude from stellar effective temperature $T_{\rm eff}$, surface gravity $\log{g}$ and metallicity (optional). \cite{andrae2022} pointed out that the metallicity results of Gaia/DR3 have a large systematic uncertainty and thus are not recommended. Therefore, we choose a function of only $T_{\rm eff}$ and $\log{g}$ as following:
\begin{equation}
    \begin{aligned}
        BC = \beta_0 + \beta_{1}x_1 +\beta_{2}x^2_1 +\beta_{3}x^3_1 +\beta_{4}x_2 +\beta_{5}x^2_2
    \end{aligned}
\end{equation}
where $x_1= \log_{10}{T_{\rm eff}/T_{\rm eff,0}}$ with $T_{\rm eff,0}=10000K$, $x_2=\log{g}$, and $\beta_0, \beta_1, \cdots, \beta_5$ are constants. For the $\GBp$ band, $\beta_0, \beta_1, \cdots, \beta_5$  equal to -0.3021, -5.1276, -0.1952, 0.0000 and 0.0836 respectively.

The apparent bolometric magnitude $m_{\rm bol}$ is then calculated from $\GBp$ taking the above BC and the interstellar extinction $A_{\GBp}$ into account:
\begin{equation}
    \begin{aligned}
        m_{\rm bol}=\GBp + BC - A_{\GBp}
    \end{aligned}
\end{equation}
which is converted to the luminosity with the distance.

The mass-luminosity relation of \citet{eker2015} is used to calculate the mass from the luminosity. \citet{eker2015} divided the stellar mass into four groups, namely low-mass, intermediate-mass, high mass and very high mass with $M/M_{\odot}$ in the range of  $0.38-1.05$, $1.05-2.40$, and $2.4-7.0$, $7.0-32$ respectively. Since the OB stars are massive, only the mass-luminosity relations for high-mass and very high-mass are adopted as follows:

\begin{equation}
    \begin{aligned}
        &\log_{10}{L}=3.962 \log_{10}{M}-0.120\  (2.4\leq M/M_\odot\le7)\\
        &\log_{10}{L}=2.726 \log_{10}{M}+1.237\  (7\leq M/M_\odot\le32)
    \end{aligned}
\end{equation}

The preliminary mass calculation eliminated 106 stars with $M<2.4{M_\odot}$. Because the mass-luminosity relation of \cite{eker2015} is based on mass rather than luminosity, the calculation is iterated. First, the mass is calculated from the luminosity by the relation for high-mass stars, and for the stars whose calculated masses exceed the range of high mass, i.e. $>7M_\odot$, the mass is re-calculated by the mass-luminosity relation for the very high mass stars. Due to the gap in luminosity between the high-mass and the very high-mass relations, 560 stars obey neither of the two relationships. However, since neither the mass nor the difference calculated by the two relations has much influence on the radio flux, we directly used the relation of very high mass stars for these 560 sources to calculate their masses.

\begin{figure}
\centering
    \subfigure[]{\includegraphics[width=8cm]{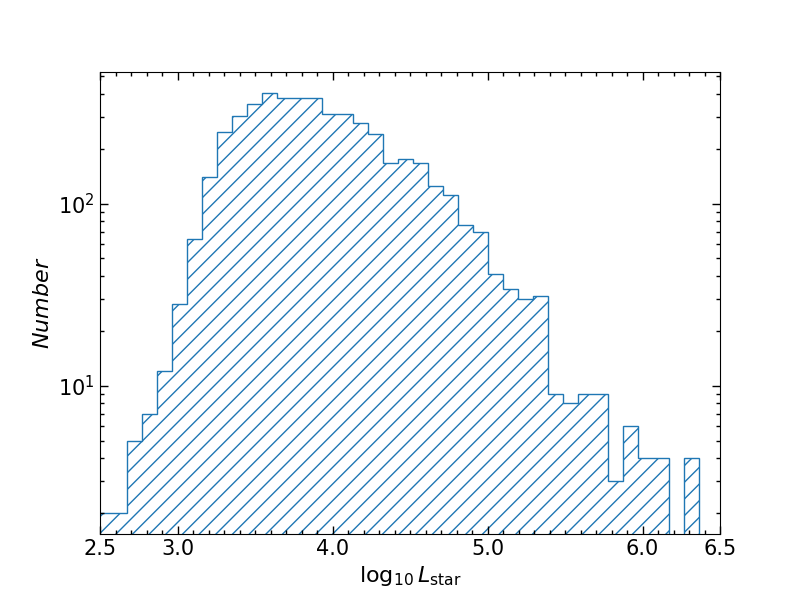}}
    \subfigure[]{\includegraphics[width=8cm]{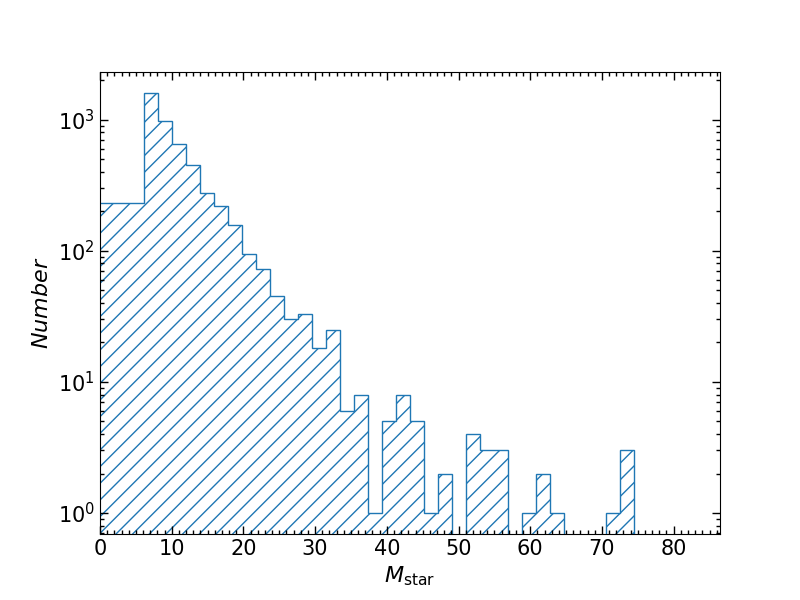}}
    \caption{Histogram of luminosity and mass of the sample stars}
    \label{fig:04}
\end{figure}

The stars whose calculated mass greater than 32$M_\odot$ are neither rejected. Although the mass of these sources may be greatly overestimated, they are definitely more massive than 30$M_\odot$, with the highest one close to 90$M_\odot$. This level of overestimation is significantly diluted in our mass loss rate equation and will bring about the difference less than an order of magnitude. Moreover, such stars are supposed to be the most luminous sources with the highest radio flux. The histograms of the luminosity and mass are shown in Figure \ref{fig:04}.

\subsection{Mass Loss Rate}

Mass loss rate ($\dot{M}$) is a key parameter to estimate the radio flux. We calculate $\dot{M}$ with the formula developed by \citet{vink2000}, which used the Monte Carlo method to fit the mass loss rate of OB-type stars as the function of stellar luminosity, mass, and effective temperature. The core idea of this method is that the total energy of the radiation is related to the total momentum gained by the outflowing matter, and the momentum in the stellar wind can be calculated from the action history of the large number of photons released below the photosphere. The specific calculation method of mass loss rate is divided into the high temperature and the low temperature groups due to the "bi-stability jump" effect around 25,000 K \citep{yu2021}.


In the higher temperature range of $27500K-50000K$, $\dot{M}$ in units of ${M_\odot/yr}$ is given by:
\begin{equation} \label{3}
    \begin{aligned}
        \log_{10}{\dot{M}}=-&6.697
        +2.194\log_{10}{(L_\star/10^5L_\odot)}
        -1.313\log_{10}{(M_\star/30M_\odot)}
        -1.226\log_{10}{(\frac{v_{\rm \infty}/v_{\rm esc}}{2.0})} \\
        +&0.933\log_{10}{(T_{\rm eff}/40000K)}
        -10.92{\log_{10}{(T_{\rm eff}/40000K)}}^2
    \end{aligned}
\end{equation}

In the lower temperature range of $12500K-22500K$, it is:
\begin{equation}\label{equ:5}
    \begin{aligned}
        \log_{10}{\dot{M}}=&-6.688
        +2.210\log_{10}{(L_\star/10^5_\odot)}
        -1.339\log_{10}{(M_\star/30M_\odot)}
        -1.601\log_{10}{(\frac{v_{\rm \infty}/v_{\rm esc}}{2.0})}
        &+1.07\log_{10}{(T_{\rm eff}/20000K)}\\
    \end{aligned}
\end{equation}

According to the work of \citet{lamers1995}, the value of the ratio $v_\infty/v_{esc}$ is 2.6 for the higher temperature and 1.3 for the lower temperature. It can be seen that the lowest temperature in Eq.\ref{equ:5} is 12500 K, which implies that stars with lower temperatures would have an insignificant mass loss rate. Therefore, we further remove the stars with $T_{\rm eff}<12500$K, which leads to the final sample of 4930 stars with mass loss rate and radio flux calculated, which is about a quarter of the initial sample.

It can be seen that the formulae lack the effective temperature range between 22500 K and 27500 K. It is due to the existence of the bi-stability jump effect, and no reliable algorithm is available for calculating $\dot{\rm M}$ in this $\Teff$ range in \citet{vink2000}. Nevertheless, since about $2\%$ of the sample stars are in this range, we still estimate their $\dot{\rm M}$  by a simple linear interpolation between the two sides of the high and low temperature ends as following:
\begin{equation}
    \begin{aligned}
    \dot{\rm M}=-3.832\times10^{-13}T_{\rm eff}+1.966*10^{-8}
    \end{aligned}
\end{equation}
This calculation may be not accurate, but it is tolerable as $\dot{\rm M}$ itself bears large uncertainty.
The current typical mass loss rate algorithms have two to three factors uncertainty \citep{matthews2018}.

\begin{figure}
    \centering
    \includegraphics[scale=0.6]{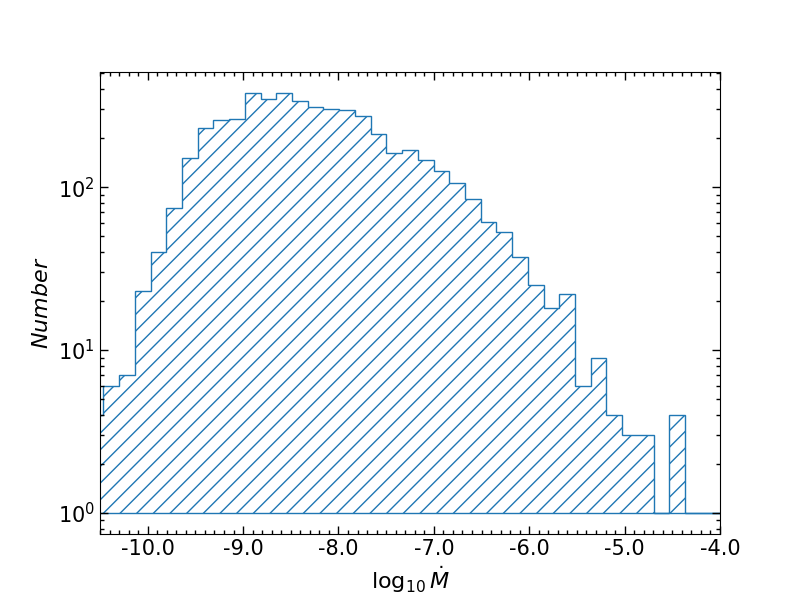}
    \caption{Histogram of the mass loss rate.}
    \label{fig:05}
\end{figure}

The distribution of the yielded $\dot{M}$ is shown in Figure \ref{fig:05}. The peak mass loss rate is around $10^{-9} - 10^{-8} {\rm M_\odot/yr}$, which is three orders of magnitude higher than $10^{-12} - 10^{-11}{\rm M_\odot/yr}$ derived in \citet{yu2021}. This difference is mainly caused by different temperature distributions. In \citet{yu2021}'s calculations, the temperature of the source simulated using the Basancon model was concentrated around 10000 K to 12500 K, while the sources in this work are mostly above 12500 K. The higher temperature implies higher luminosity and higher $\dot{M}$.  There are 97 sources with $\dot{M}>10^{-6} {M_\odot/yr}$.

\section{Radio flux} \label{sec:Radio flux}

\subsection{The result} \label{subsec:result}

According to the classical spectral line-driven stellar wind theory, the radio emission of OB stars originates from the thermal free-free radiation of ionized gas around the star resulting from mass loss \citep{wright1975, lamers1999, vink2011}. The flux density $S_\nu$ in units of mJy at a frequency $\nu$ in GHz in the radio band is related to the mass loss rate $\dot{\rm M}$ in ${\rm M_\odot yr^{-1}}$ under the assumption of spherically symmetric uniform wind as follows: \citep{wright1975, panagia1975}
\begin{equation} \label{eq:Sv}
    S_\nu=\frac{2.30\times10^{7}\dot{M}^{4/3}Z^{4/3}(\gamma g_{\rm ff}\nu)^{2/3}}{\mu^{4/3} v_\infty^{4/3} D^{2}}
\end{equation}
where $\mu$ is the average ion weight, $v_\infty$  the terminal velocity in km/s, $D$  the distance in kpc, $Z$ the effective charge per ion, $\gamma$  the average number of electrons per ion, and $g_{\rm ff}$ is the free-free Gunter factor given by:
\begin{equation}
    g_{\rm ff}=-1.66+1.27\lg(T_{\rm wind}^{3/2}/(Z\nu))
\end{equation}
with $T_{\rm wind}$ being the temperature of the radio photosphere layer.

We assume that the stellar wind of OB stars is completely ionized, and we take the cosmic abundance of elements which gives $Z$, $\gamma$, and $\mu$ values of 1, 1, and 1.26 respectively. The observation frequency is set to be 1450MHz, the terminal velocity is 1000km/s, and $T_{\rm wind}$ is set to be 0.5 $T_{\rm eff}$ \citep{lamers1993, abbott1981}. The flux at other frequency can be calculated from $S_{\rm 1450MHz}$ by the relation of $S_\nu \propto \nu^{\rm 0.67}$.

The histogram of the calculated flux is displayed in Figure \ref{fig:06}. The peak of the distribution is at about $10^{\rm -8}$Jy. In comparison, this value is $10^{\rm -13}$Jy in \citet{yu2021},i.e. five orders of magnitude lower. The above calculation indicates that the peak of mass loss rate distribution in the present work is three orders of magnitude higher and would lead to a higher radio flux by four orders of magnitude. The other order can be attributed to the distance with its peak being at about 2 kpc for the present sample, a few factors closer than that ($\sim$ 5kpc) in \citet{yu2021}. Moreover, the Besancon model that \citet{yu2021} used is a virtual model that contains no real stars. It is a model for the galaxy that is populated by creating stars so that it fits the expected distribution. This also explains the difference with \citet{yu2021}; the Besancon model does not include clusters and, therefore, misses the clustering of the most massive stars, but is dominated by the less massive yet much more common stars.

\begin{figure}
    \centering
    \includegraphics[scale=0.4]{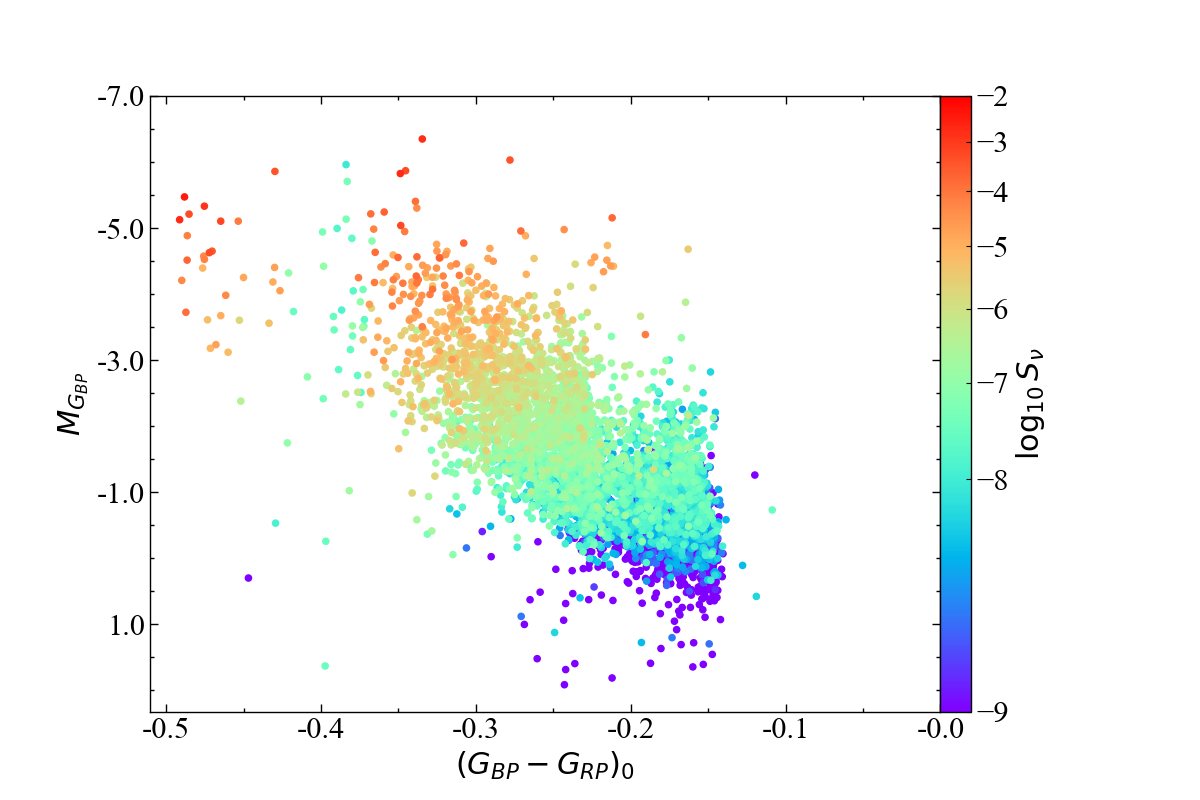}
    \caption{The distribution of the calculated $S_{\rm 1.4GHz}$ in the color-magnitude diagram  of the  sample stars}
    \label{fig:07}
\end{figure}

It is worth mentioning that, according to \citet{vink2000}, there are two peaks of the mass loss rate of stars at $\sim$20,000\,K and $\sim$40,000\,K respectively. In the range from 22500\,K to 27500\,K, stellar mass loss rate decreases rapidly with increasing temperature, and at around 27500 K, the mass loss rate of stars is similar to that of stars at around 12500 K. Therefore, we can clearly see in Figure \ref{fig:07} that there are a number of sources with very small $S_\nu$, at the position of $(G_{\rm BP}-G_{\rm RP})_0$ = -0.37 to -0.4 that corresponds to the  temperature between 22500\,K and 27500\,K. The sources with very high $S_\nu$ are concentrated between $(G_{\rm BP}-G_{\rm RP})_0 = -0.35$ and $ -0.5$, corresponding to temperatures of $\sim$20,000K and $\sim$40,000K.

\subsection{The spatial distribution}

The distribution of the sources is displayed in Figure \ref{fig:all sky map} with the radio flux denoted by color-bar. As expected due to their young age, massive stars are mostly distributed near the Galactic plane, with high flux sources concentrated within the plane. However, there are some outlying sources beyond the Galactic plane, although they have relatively low fluxes. It should be noted that the radius of the circle in the figure is not proportional to the size of the actual sky area of the OB association. When comparing with the nearby OB associations assembled by \citet{ruprecht1966}, a few are apparently matched, such as Mon OB2, Gem OB2 and Cyg OB2, meanwhile, Ori OB2 and Cep OB2 are not well matched. Furthermore, a few OB associations are beyond the survey area covered by LAMOST. On the other hand, some clusters of OB stars were not in the early list of OB associations. This mis-match may be caused by the incomplete sky-coverage of the LAMOST survey.

\subsection{The sample detectable by FAST and SKA}

\begin{figure}
    \centering
    \includegraphics[scale=0.6]{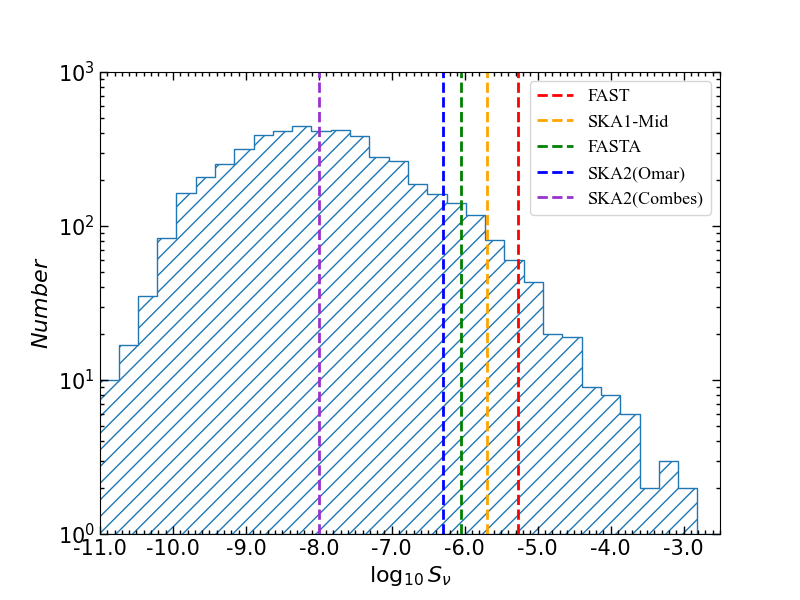}
    \caption{Histogram of the calculated $S_{\rm 1.4GHz}$, where the sensitivity of FAST/FASTA/SKA is marked by dashed line. The two sensitivities of SKA2 come from \citet{omar2023} and \citet{combes2015} respectively.}
    \label{fig:06}
\end{figure}

The sensitivity of FAST is calculated at the frequency of 1450MHz with a system temperature of 30\,K (\cite{jiang2020}). With an on-source integration time of 30-min and a bandwidth of 200MHz, the sensitivity is approximately 10 $\mu$Jy at a 5-sigma level. As shown in Figure \ref{fig:06}, the flux of 82 stars, or 1.62$\%$ of the total sample, is above this level. On the other hand, the very high sensitivity leads to confusion that is about 1mJy for the continuum observation. By taking the confusion limit into account, only five stars are detectable. However, the uncertainty in the estimated radio flux is significant, originated from the uncertain mass loss rate as well as the model of the wind. Therefore, the brightest OB stars are still potential objects with the thermal wind emission detectable by FAST. In Table 1, we listed the brightest 13 stars with $S_\nu > 0.1$mJy in the sample which deserves a try by FAST.

The estimated observing power of SKA1-mid is based on the performance given by \citet{braun2019}, who calculated that its point source sensitivity would be 2$\mu$Jy within an integration time of 1 hour. For the sensitivity of SKA2, \citet{braun2019} stated that it is still uncertain. However, from the overall telescope configuration, the performance of SKA2 is about 20 times that of SKA1. Therefore, \citet{omar2023} cited \citet{braun2019} and gave 0.5$\mu$Jy as a reference value. We also use this value as the sensitivity of SKA2 for comparison.With this sensitivity, there would be 392 stars detectable by SKA1-mid and 749 stars detectable by SKA2.
In Figure \ref{fig:06}, we also plotted the estimated maximum observational capability of SKA2 by \citet{combes2015} as a reference cited by \citet{yu2021} who calculated that the point source sensitivity would be close to 10 nano-Jy within an integration time of 8 hours.

\begin{figure}
    \centering
    \includegraphics[scale=0.7]{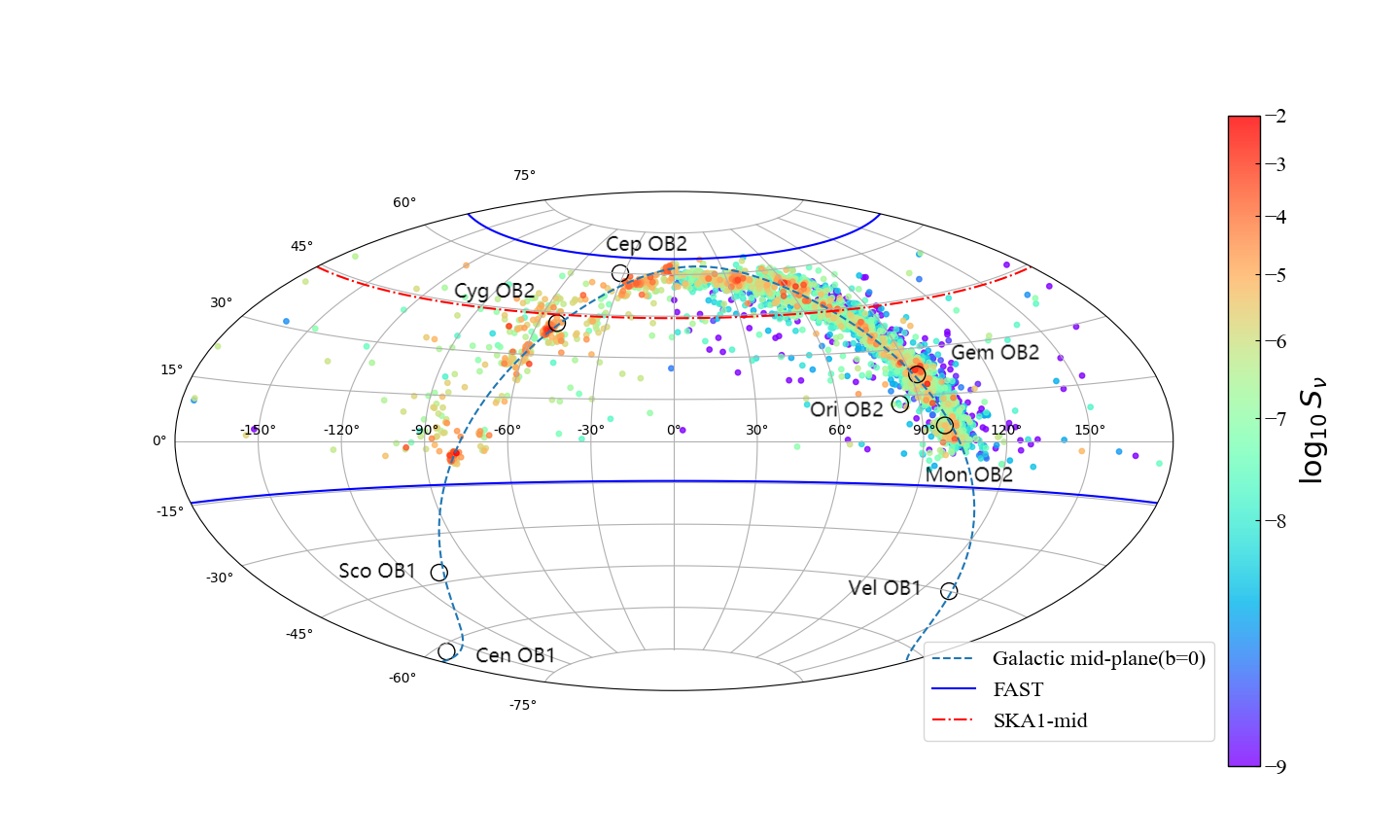}
    \caption{The spatial distribution of the sample stars in the equatorial coordinate, where the Galactic plane is denoted by blue dashed line, and the latitude limit of FAST and SKA1-mid is denoted by blue solid line and red dashed line respectively. Several famous OB associations are labelled.} 
    \label{fig:all sky map}
\end{figure}

The observable sky area of FAST and SKA1-mid is marked as well in Figure \ref{fig:all sky map}. The geographical latitude of FAST telescope is 25.6\degr  north and its meridian-like structure  can move within the zenith angle less than 40\degr, so its observable range is between -14.4\degr and 65.6\degr in declination which is marked by the blue line in Figure \ref{fig:all sky map}. Because the LAMOST telescope locates at a latitude of 40\degr, nearly all the sources are within the FAST observable area. On the other hand, SKA will be located in the southern hemisphere. If we take an elevation of 15\degr above the horizon as the limit, its observable sky area is south of 44.3\degr\,N in declination, which is marked by the red line in Figure \ref{fig:all sky map}. Due to the large difference in geographical latitude of SKA with LAMOST, many of the candidate sources are beyond the SKA sky. Taking  the  SKA's location into account, only part of the objects is visible to SKA1-mid, which reduces the number of detectable stars from 392 to 174.

A plan in China to build a FAST array (FASTA) is under discussion. For example, an array of six 500-meter telescope with more advanced receivers would not only increase the sensitivity by nearly an order of magnitude, but also alleviate the problem of confusion of a single dish. Such facility will be very powerful to detect radio emission of stars. In the case of OB stars, the number of stars detectable will be 583, which is several tens times of the present performance of FAST.

\section{Discussion}

\subsection{The Cyg OB2 association} \label{sec:CygOB2}

The e-Merlin Cyg OB2 Radio Survey (COBRaS) project has observed 1000 OB stars in the Cygnus OB2 association for a total of about 300 h at 5 GHz and 1.6 GHz \citep{willis2011}.
According to \citet{morford2015}, two O stars (Cyg OB2 \#7 and A15) and one B star (Cyg OB2 \#12) have model-predicted flux $S_{\rm 1.4GHz}$ of 20.5, 16.7, and 1770 $\mu Jy$ respectively, but only Cyg OB2 \#12 is detected with $S_{\rm 1.4GHz} = 341 {\rm \mu Jy}$, significantly lower than the predicted. Meanwhile, the two O stars are not detected and have an upper limit of 150 and 138${\rm \mu Jy}$ respectively. By detailed study of four O stars and five B stars observed in this project, \citet{morford2016} inferred the upper limit of mass loss rate from the non-detection at 1.4GHz for four O stars and four B stars to be smaller than $5\times10^{-6} {\rm M_{\odot}/yr}$, which is in general agreement with the model within one order of magnitude. For Cyg OB2 \#12, the only detected source,  they derived $\dot{M}$ to be $5.4\times10^{-6} {\rm M_{\odot}/yr}$ from $S_{\rm 1.4GHz} = 1013 {\rm \mu Jy}$, which is about fifth of the model predicted $24.5\times10^{-6} {\rm M_{\odot}/yr}$. The multi-epoch observations indicate that the radio flux of Cyg OB2 \#12 is variable, and there is significant discrepancy between the observation and the model prediction.

In comparison with our predicted mass loss rate, \citet{morford2016} used the same mass loss rate model \citep{vink2000}. However, \citet{morford2016} used the specific values of stellar luminosity and mass for its nine observation targets cited from \citet{clark2012} and other five papers, which are different from our calculations. For the nine stars in \citet{morford2016}, most of them in our results have much lower effective temperatures. Since these stars are mostly in the main-sequence, the lower effective temperature leads to smaller luminosity, mass and mass loss rate.
On the other hand, for stars whose effective temperature and surface gravity are close to the observed targets of \citet{morford2016}, their masses exceed 32 solar masses in our estimation. Due to our mass algorithm, the estimated mass of these stars may be overestimated already. When comparing with the observation results of \citet{morford2016}, this overall impact means that there may be an overestimation of the radio flux of stars with the largest mass.

According to \citet{knodlseder2000}, Cyg OB2 is centered at about ${\rm RA=20^h33^m10^s}$ and  ${\rm DEC=+41^\circ12^\prime}$ (J2000) with an angular diameter of about $2^\circ$. \citet{berlanas2019} showed that the Cyg OB2 has two structures at a distance of $\sim$1.76kpc and $\sim$1.35kpc respectively.  Fourteen stars in our sample are located within this area and listed in Table \ref{tab:Cyg OB2 stars}, two of which are very possibly beyond the distance range of the association. Unfortunately, the members mentioned above are not included in the sample. Instead, the sample stars are generally much fainter and the estimated $S_{\rm 1.4GHz}$ is on the order of micro-Jy.

\subsection{Cross-match with the NVSS catalog} \label{sec:NVSS}

The sample stars are cross-matched with  the NRAO VLA Sky Survey (NVSS) Catalog \citep{condon1998}. The NVSS Catalog covers the sky north of the declination of -40\degr\ at 1.4 GHz, identical with the area of the sample and the frequency in calculation. It includes a catalog of almost two million discrete sources brighter than a flux density of about 2.5 mJy. The rms uncertainties in right ascension and declination are $\sim 7\arcsec $ at the survey limit that is the expected value of the OB stars. By taking a radius of $15\arcsec $ for cross-identification, a total of 13 sources are found.

These thirteen objects are queried through the SIMBAD database for their classification. It is found that they belong to three categories: seven confirmed stellar objects, three confirmed non-stellar objects, and three unidentified, which are listed in Table \ref{tab:NVSS}. Their estimated fluxes in this work are generally several orders of magnitude below the NVSS sensitivity. It seems they are false matches. Therefore, a further examination for all confirmed stellar objects is performed. The SIMBAD data is searched within a 45 arc-second range near each source and the results are present in Table \ref{tab:NVSS}. Within the 45 arc-second area, four sources are marked as A because there is no objects, and one source as B because there are other stars or uncertain data that could influence the result, and two sources as C because there are clear radio source(s). It can be assumed that the cross identification is unreliable if there are other sources within this range, i.e. the Class B and C sources.

\subsection{The influence of binary}

Early-type stars have a high binary rate. In early-type binary stars, if both stars are OB stars or WR stars, their strong stellar winds will collide and produce extremely strong non-thermal synchrotron radiation. This non-thermal synchrotron radiation is usually considered to be produced by relativistic particles. Since the radio photosphere of these early-type stars is about hundreds of stellar radii, if the binary star is wide (period  greater than 10 years), then the non-thermal synchrotron radiation will be produced outside the radio photosphere of the star. If the period  of the binary star is less than 1 year, then the non-thermal synchrotron radiation from the wind collision region will not be observable because it will be absorbed in the optically thick region of the individual star's wind \citep{eichler1993,van2008,de2013,de2017}.

\citet{luo2021} used the data from LAMOST/DR5 to derive a binary probability of $0.4^{+0.05}_{-0.06}$ for OB stars.  Here, we take the selection method used by \citet{guo2022} to find the binaries in our sample. \citet{guo2022} calculated the radial velocity of 9382 early-type stars in the LAMOST medium-resolution survey data in the \citet{guo2021} table and identified binaries according to the velocity dispersion with the following criteria

\begin{equation} \label{eqRV}
    \begin{gathered}
        \frac{\left| v_i-v_j \right|}{(\sigma_i^2+\sigma_j^2)^{\frac{1}{2}}} > 4 \\ \left| v_i-v_j \right| > C
    \end{gathered}
\end{equation}
\\
where $v_i$ and $\sigma_i$ are the radial velocity and the associated uncertainty measured for the spectrum at epoch i(j) respectively. In order to eliminate pulsating variables, the constant C is set to 15.57 km/s \citep{guo2022}.

According to this criteria, 95 of the 559 objects with the radial velocity measurements in our sample are binaries, i.e about 17\%. This fraction is smaller than 40\% derived by \citet{luo2021}.  Due to the lack of specific parameters about the binary orbits, we cannot further determine the binary contribution to the radio flux. However, it is certain that the radio flux of some binaries would be increased by the wind interaction. On the other hand, the measurement of radio flux may become an indicator of binary.
Based on \citet{van2008}'s observation of Cyg OB2 No.9's radio flux at 6cm wavelength using VLA, they thought that the thermal synchrotron radiation of this binary was 2mJy, and the non-thermal synchrotron radiation varied between 0-6mJy with a period of 2.35 years. Taking this as reference, we estimate that if the binaries in our data are double early-type stars, their non-thermal synchrotron radiation will increase their radio flux by up to three times. If their companions are not OB stars or WR stars with high mass loss rates, then their radio flux will remain unchanged.

\subsection{The effects of clumping}

The stellar wind of OB stars exhibits dynamic instability, which leads to the formation of shocks and clumps. These clumps enhance radio emission and cause the spectral index $\alpha$ to vary with frequency. \citet{williams1990} found a significant deviation from $\alpha$ = 0.6 at high frequencies in their study of Wolf-Rayet star $\gamma$ Velorum at submillimeter and infrared frequencies. \citet{nugis1998} found that the winds of WN and WC stars exhibit steepening phenomena in the spectrum with $\alpha$ = 0.77 and 0.75, respectively, deviating from $\alpha$ = 0.6 due to clumping of stellar wind material.

\citet{daley2016} used ALMA to study the effect of clumping on spectral index $\alpha$ within the range of $10$GHz to $10^4$GHz. The results showed that clumping would cause spectral index $\alpha$ to rise from the standard 0.6 to between 0.7-0.8 when the frequency is below 500GHz.The spectral index $\alpha$ used in our radio flux calculation formula was 2/3 (0.66). According to \citet{daley2016}'s results, we take 0.6 and 0.8 as the minimum and maximum values of spectral index $\alpha$ under clumping effect.
It is evident from Eq.\ref{eq:Sv} that the influence of spectral index variations in the range of 0.6 to 0.8 on the radio flux $S_{\nu}$ is weak. A change in $\alpha$ from 0.66 to 0.6 results in a decrease of $2.3\%$ in the flux, while a change in $\alpha$ from 0.66 to 0.8 will result in an increase of $4.5\%$ in the flux. Even if we take the extreme case that $\alpha = 2.0$, the flux would be increased by 57\%, still within a factor of two.

\section{Summary}

Based on the \cite{guo2021} OB star catalog and the newly published Gaia/DR3 stellar parameters, we quantitatively calculated and analyzed the radio flux of OB stars in the catalog to estimate whether they are likely to be detected by FAST and SKA. We conducted further screening of the OB star candidates listed in \cite{guo2021}, and separated out the hot sub-dwarfs, main sequence OB stars and giant stars through the color-magnitude diagram. The mass loss rate is derived from stellar parameters, which are then used to estimate the apparent flux at 1450MHz under the assumption of free-free emission mechanism with stellar distance.

The mass loss rate of about 4930 OB stars range from about $10^{-4}$ to $10^{-11} {\rm M_{\odot}/yr}$, and correspondingly, the flux at 1.4GHz ranges from $10^{-3}$ to $10^{-11} {\rm Jy}$. The high flux stars are distributed mostly in the Galactic plane as expected for massive stars. By comparing with the sensitivity and considering the sky coverage, 82 objects are detectable by FAST if confusion is neglected. But the number would be greatly reduced to five if confusion is taken into consideration. So a FAST array would be extremely helpful to detect such point-like stellar sources and important to understand stellar wind and evolution. For the SKA, there will be 392 and 749 stars detectable by SKA1-mid and SKA2 respectively.

We estimated the binary proportion in our source table. Ninety-five out of the 559 objects in our sample with radial velocity measurements are binaries, i.e., about 17$\%$. Assuming the extreme case that the binary probability for the entire population is higher, at 40$\%$, and that all binaries are double early-type stars with triple the flux, the number of observable sources for FAST, SKA1-mid, and SKA2 would increase to 124, 504, and 919, respectively, with an increase of 42, 112, and 170 sources.

\section{acknowledgments}
We are grateful to Drs Mengyao Xue, Yanjun Guo and Jing Tang for their helpful discussion. The authors are grateful to the anonymous referee whose comments helped to improve the work significantly.  This work is supported by the NSFC project 12133002, National Key R\&D Program of China No. 2019YFA0405503, and CMS-CSST-2021-A09. This work has made use of the data from LAMOST and Gaia.

\bibliographystyle{aasjournal}
\bibliography{Estimation_of_the_flux_at_1450MHz_of_OB_stars_for_FAST_and_SKA.bib}


%
%


\begin{table}[]
 \caption{The $S_{\rm 1.4GHz}$ brightest OB stars}
    \centering
    \begin{tabular}{c c c c c c c c c c c c c}
    \hline \hline
        RA & DEC & Distance & $G_{\rm BP}$ & $G_{\rm RP}$ & $A_{G_{\rm BP}}$ & $M_{G_{\rm BP}}$ & $C^0_{G_{\rm BP},G_{\rm RP}}$ & $T_{\rm eff}$ & $L_{\rm star}$ & $M_{\rm star}$ & $\dot{M}$ & $Flux$\\
          ($^{\circ}$) & ($^{\circ}$) & pc & & & & & & $10^4 {\rm K}$ & $10^5 {\rm L_{\odot}}$ & ${\rm M_{\odot}}$ & $10^{-6} {\rm M_{\odot}/yr}$ & ${\rm mJy}$ \\
        \hline
55.1326 & 50.7481 & 2596 & 11.6 & 10.27 & 4.0 & -4.65 & -0.47 & 3.41 & 9.9 & 56 & 7.5 & 0.156\\
101.2242 & 0.6202 & 4109 & 9.34 & 8.67 & 2.08 & -6.03 & -0.28 & 1.74 & 8.0 & 51 & 16.9 & 0.172 \\
69.2314 & 53.876 & 3966 & 11.33 & 10.0 & 3.9 & -5.86 & -0.43 & 2.96 & 19.8 & 72 & 16.0 & 0.181\\
306.7926 & 38.9004 & 1732 & 10.92 & 9.48 & 4.27 & -4.63 & -0.47 & 3.21 & 9.2 & 54 & 5.6 & 0.237 \\
94.8142 & 12.1818 & 6203 & 12.63 & 11.5 & 3.57 & -5.21 & -0.49 & 4.09 & 21.6 & 74 & 38.3 & 0.244\\
95.0133 & 24.2293 & 4218 & 10.47 & 9.5 & 2.91 & -5.87 & -0.35 & 2.01 & 8.7 & 53 & 22.8 & 0.248 \\
88.1402 & 25.7641 & 3085 & 11.56 & 10.22 & 4.03 & -5.1 & -0.46 & 3.4 & 14.6 & 64 & 14.5 & 0.266\\
36.0093 & 57.3539 & 2154 & 9.28 & 8.48 & 2.53 & -5.04 & -0.35 & 2.05 & 5.0 & 43 & 8.9 & 0.271 \\
64.5241 & 53.6187 & 3847 & 12.39 & 10.82 & 4.57 & -5.33 & -0.48 & 3.75 & 21.8 & 74 & 35.2 & 0.563 \\
60.2266 & 55.2278 & 3757 & 10.52 & 9.19 & 3.72 & -6.35 & -0.33 & 1.95 & 13.1 & 62 & 44.5 & 0.760\\
92.4183 & 23.0729 & 3302 & 10.58 & 9.74 & 2.9 & -5.13 & -0.49 & 4.05 & 21.5 & 74 & 37.6 & 0.840\\
281.3574 & -4.0169 & 2818 & 11.37 & 9.63 & 4.72 & -5.83 & -0.35 & 2.14 & 10.2 & 56 & 32.2 & 0.884\\
94.9165 & 18.3715 & 3423 & 10.58 & 9.64 & 3.11 & -5.47 & -0.49 & 4.1 & 28.8 & 82 & 62.9 & 1.550 \\
    \hline
    \end{tabular}
    \label{tab:Brightest stars}
\end{table}

\begin{table}[]
 \caption{OB stars in the Cyg OB2 range}
    \centering
    \begin{tabular}{c c c c c c c c c c c c c}
    \hline \hline
        RA & DEC & Distance & $G_{\rm BP}$ & $G_{\rm RP}$ & $A_{G_{\rm BP}}$ & $M_{G_{\rm BP}}$ & $C^0_{G_{\rm BP},G_{\rm RP}}$ & $T_{\rm eff}$ & $L_{\rm star}$ & $M_{\rm star}$ & $\dot{M}$ & Flux\\
          ($^{\circ}$) & ($^{\circ}$) & pc & & & & & & $10^4 {\rm K}$ & $10^4 {\rm L_{\odot}}$ & ${\rm M_{\odot}}$ & $10^{-9} {\rm M_{\odot}/yr}$ & $10^{-3}{\rm mJy}$\\
        \hline
306.7644 & 40.0505 & 977 & 11.26 & 10.7 & 1.65 & -0.42 & -0.19 & 1.39 & 0.3 & 7 & 1.1 & 0.016 \\
307.7038 & 40.7569 & 1295 & 12.57 & 11.61 & 2.67 & -0.69 & -0.24 & 1.5 & 0.5 & 8 & 2.5 & 0.029 \\
308.4219 & 41.5515 & 1242 & 11.66 & 10.86 & 2.19 & -1.13 & -0.18 & 1.34 & 0.6 & 8 & 2.5 & 0.031 \\
307.3004 & 40.5779 & 1697 & 12.55 & 11.26 & 3.31 & -2.0 & -0.18 & 1.25 & 1.1 & 11 & 7.8 & 0.074 \\
305.7393 & 40.7336 & 1677 & 13.58 & 12.06 & 4.0 & -1.57 & -0.24 & 1.51 & 1.2 & 11 & 10.4 & 0.114 \\
309.7769 & 41.6833 & 1347 & 13.95 & 12.12 & 4.75 & -1.51 & -0.25 & 1.5 & 1.1 & 11 & 8.7 & 0.139 \\
308.5251 & 40.718 & 1243 & 12.76 & 11.27 & 3.91 & -1.66 & -0.24 & 1.5 & 1.3 & 11 & 11.6 & 0.239 \\
307.1527 & 40.6245 & 1539 & 12.03 & 10.82 & 3.2 & -2.24 & -0.22 & 1.42 & 1.8 & 13 & 19.1 & 0.303 \\
310.1077 & 41.6242 & 1277 & 13.67 & 11.58 & 5.35 & -2.24 & -0.24 & 1.5 & 2.2 & 14 & 29.1 & 0.776 \\
306.9336 & 41.5497 & 1284 & 9.92 & 9.64 & 1.2 & -1.87 & -0.28 & 1.74 & 2.1 & 13 & 31.5 & 0.867 \\
307.3386 & 39.9884 & 1647 & 12.44 & 10.95 & 3.9 & -2.66 & -0.24 & 1.53 & 3.1 & 16 & 54.2 & 1.071 \\
306.9359 & 40.4274 & 1732 & 12.85 & 11.15 & 4.4 & -2.79 & -0.24 & 1.5 & 3.5 & 16 & 66.9 & 1.28 \\
308.0163 & 39.8207 & 1330 & 16.04 & 12.68 & 8.44 & -3.17 & -0.23 & 1.5 & 4.5 & 18 & 103.5 & 3.885 \\
306.1209 & 40.6174 & 2208 & 12.65 & 10.93 & 4.43 & -3.61 & -0.24 & 1.5 & 7.1 & 21 & 222.8 & 3.917 \\

    \hline
    \end{tabular}
    \label{tab:Cyg OB2 stars}
\end{table}

\begin{table}[]
 \caption{
 The objects cross-matched with the NVSS catalog by a radius of $15\arcsec$. They are divided into three classes by the SIMBAD database:  the first part as stellar sources, the second part as non-stellar sources, and  the third part as unidentified. The stellar sources are further divided into three groups according to the situation within a 45 arc-second circle: marked as A  if there is no objects; marked as B if there are other stars,  and marked as C if there are other radio sources.}
    \centering
    \begin{tabular}{c c c c c c c c c c c c}
    \hline \hline
    Source name & $\sigma_{\rm RA}^{\rm NVSS} $ & $\sigma_{\rm Dec}^{\rm NVSS}$ & Distance & $T_{\rm eff}$ & ${M_{\rm star}}$ & ${L_{\rm star}}$ & $\dot{M}$ & $S_\nu^{\rm NVSS}$ & $S_\nu^{\rm this work}$ & $Flag^{45\arcsec}_{SIMBAD}$\\
    & (\arcsec) & (\arcsec) & $(pc)$ & $10^4 {\rm K}$ &${M_\odot}$ & $10^3 {\rm L_\odot}$ & $10^{-10}{M_\odot/yr}$ & mJy  & mJy & \\
    \hline
\textit{stellar sources}\\
TYC 2440-341-1 & 0.05 & 0.7 & 1718 & 1.3 & 5 & 1.1 & 1.2 & 29.8 & $2.8\cdot 10^{-7}$ & A\\
SDSS J062231.18+262932.4 & 0.3 & 3.6 & 3996 & 1.4 & 6 & 2.1 & 4.3 & 4.1 & $2.8\cdot 10^{-7}$ & A\\
TYC 3661-1648-1 & 0.2 & 1.8 & 1720 & 1.3 & 7 & 3.6 & 11.7 & 9.5 & $5.8\cdot 10^{-6}$ & A\\
Gaia DR3 442249316436001152 & 0.25 & 2.4 & 1909 & 1.5 & 10 & 8.1 & 53.8 & 6.2 & $3.7\cdot 10^{-5}$ & B\\
TYC 154-1153-1 & 0.04 & 0.6 & 1665 & .6 & 9 & 7.8 & 54.3 & 53.9 & $4.9\cdot 10^{-5}$ & C\\
MFJ SH 2-272 3 & 0.13 & 1.6 & 3476 & 1.7 & 14 & 21.1 & 311.0 & 12.4 & $1.2\cdot 10^{-4}$ & C\\
BD+54 3 & 0.22 & 2.0 & 2784 & 2.0 & 22 & 76.6 & 3479.0 & 9.1 & $4.6\cdot 10^{-3}$ & A\\
\hline
\textit{non-stellar}\\
NVSS J055708+210742 & 0.04 & 0.6 & 2885 & 1.3 & 6 & 1.7 & 2.6 & 71.5 & $2.7\cdot 10^{-7}$ &\\
Lan 76 & 0.57 & 9.6 & 4438 & 1.6 & 16 & 34.3 & 672.5 & 2.5 & $2.0\cdot 10^{-4}$ &\\
NVSS J061453+122122 & 0.03 & 0.6 & 3254 & 1.7 & 19 & 56.7 & 1725.7 & 329.6 & $1.3\cdot 10^{-3}$ &\\
\hline
\textit{unidentified}\\
NVSS 065159+135744 & 0.17 & 2.4 & 4958 & 1.3 & 6 & 1.9 & 3.2 & 5.9 & $1.3\cdot 10^{-7}$ &\\
NVSS J062542.25+195956.0 & 0.57 & 8.9 & 3548 & 1.6 & 9 & 6.2 & 36.2 & 13.1 & $6.3\cdot 10^{-6}$ &\\
NVSS J195848+404205 & 0.04 & 0.6 & 4645 & 1.4 & 10 & 10.1 & 71.7 & 755.6 & $9.0\cdot 10^{-6}$ &\\
    \hline
    \end{tabular}
    \label{tab:NVSS}
\end{table}



\end{CJK}
\end{document}